\documentclass[aps,prb,twocolumn,superscriptaddress,showkeys,amsmath,amssymb]{revtex4-1}

\usepackage{graphicx}
\usepackage[caption=false,labelformat=parens]{subfig}
\usepackage{bbm,upgreek,mathtools}
\usepackage{dcolumn}
\newcolumntype{.}{D{.}{.}{-1}}
\usepackage{booktabs}
\usepackage{hyperref}

\begin{document}

\title{Ultrafast Molecular Transport on Carbon Surfaces:\\ The Diffusion of Ammonia on Graphite}

\author{A. Tamt\"{o}gl}
\email{tamtoegl@gmail.com}
\affiliation{Institute of Experimental Physics, Graz University of Technology, 8010 Graz, Austria}
\author{M. Sacchi}
\affiliation{Department of Chemistry, University of Surrey, GU2 7XH, Guildford, United Kingdom}
\author{I. Calvo-Almaz\'{a}n}
\affiliation{Cavendish Laboratory, University of Cambridge, J. J. Thomson Avenue, CB3 0HE, Cambridge, United Kingdom.}
\affiliation{Material Science Division, Argonne National Laboratory, Argonne, Illinois, 60439, United States}
\author{M. Zbiri}
\affiliation{Institut Laue-Langevin, 71 Avenue des Martyrs, CS 20156, F-38042 Grenoble Cedex 9, France.}
\author{M. M. Koza}
\affiliation{Institut Laue-Langevin, 71 Avenue des Martyrs, CS 20156, F-38042 Grenoble Cedex 9, France.}
\author{W. E. Ernst}
\affiliation{Institute of Experimental Physics, Graz University of Technology, 8010 Graz, Austria}
\author{P. Fouquet}
\affiliation{Institut Laue-Langevin, 71 Avenue des Martyrs, CS 20156, F-38042 Grenoble Cedex 9, France.}


\begin{abstract}
We present a combined experimental and theoretical study of the self-diffusion of ammonia on exfoliated graphite. Using neutron time-of-flight spectroscopy we are able to resolve the ultrafast diffusion process of adsorbed ammonia, NH$_3$, on graphite. Together with van der Waals corrected density functional theory calculations we show that the diffusion of NH$_3$ follows a hopping motion on a weakly corrugated potential energy surface with an activation energy of about 4 meV which is particularly low for this type of diffusive motion. The hopping motion includes further a significant number of long jumps and the diffusion constant of ammonia adsorbed on graphite is determined with $D=3.9 \cdot 10^{-8}~\mbox{m}^2 /\mbox{s}$ at 94 K.
\end{abstract}

\maketitle

\section{Introduction}
The diffusion of ammonia on graphite is particularly interesting for potential applications of graphene and graphitic material surfaces. Those include chemical doping of graphene, e.g., n-doping of graphene by thermal annealing in the presence of ammonia gas\cite{Liu2011,Wang2009}. Furthermore, the modification of the electronic structure of graphene upon adsorption of ammonia has been employed for quantum sensing / gas sensor applications\cite{Romero2009,Paul2012,Yuan2013,Aziza2014}. It was shown that it is possible to use graphene as a gas sensor with high sensitivity and high accuracy for detecting ammonia groups due to the fact that ammonia adsorbed on graphene induces the appearance of new substrate electronic states\cite{Zhang2015,Schedin2007,Boettcher2011}. The changes to the graphene electronic states could be reverted by annealing, where in particular desorption is often dominated by the kinetic processes on the surface. Moreover, the gas adsorption and diffusion on the graphene surface basically determines the sensitivity of these graphene based gas sensors\cite{Sun2017}.\\
The adsorption and diffusion of molecular species on graphene and graphitic materials is also of fundamental interest in various fields. Several studies on the dynamics and the structure of physisorbed molecular species on graphite have been carried out, including molecular hydrogen\cite{Bienfait1999}, alkanes\cite{Thomas1982,Clark2001,Arnold2002,Arnold2002b,Bruch2007,Arnold2012} and aromatic hydrocarbons\cite{Calvo2016,Calvo2014,Hedgeland2009}. The diffusion of adsorbates and clusters on carbon-based materials has also been subject to intensive research, in search for low-friction and superdiffusive systems\cite{Bardotti1995,Miura2003,Guerra2010,Kawai2016} as well as for studying elementary dynamic processes such as atomic-scale friction\cite{deWijn2011,Pawlak2017} and the development of nanometer size motorization systems\cite{Browne2006}.\\
However, little experimental data exists for the diffusion of ammonia (NH$_3$) on graphite. This is quite surprising, given that NH$_3$ represents one of the simplest heteroatomic molecules. Experimental results about the ammonia/graphite system are mainly based on thermal desorption studies of ammonia on graphitic surfaces and some very early neutron and nuclear magnetic resonance (NMR) diffusion data\cite{Tabony1980}. While ammonia on highly oriented pyrolytic graphite (HOPG) starts to desorb at 90 K\cite{Ulbricht2006}, slightly higher desorption temperatures (111 K) have been found for graphene/metal systems\cite{Boettcher2011}. According to density functional theory (DFT) calculations, NH$_3$ adsorbs in the centre of the carbon hexagon ($E_a = 31 - 48$ meV ), almost invariant to rotations around the axis perpendicular to the surface and through the nitrogen atom\cite{Zhang2015,Leenaerts2008,Lin2013}. On the other hand, the adsorption energy from thermal desorption spectroscopy (TDS) is $E_a = (260\pm20)~\mbox{meV}$\cite{Ulbricht2006} and DFT calculations have predicted that the barrier for translational diffusion is about 10 meV \cite{Zhang2015,Tabony1980}.\\
Here we present a combined neutron scattering and density functional theory (DFT) study of the diffusion of ammonia on exfoliated graphite. Scattering techniques such as quasi-elastic neutron scattering (QENS) and quasi-elastic helium atom scattering (QHAS) are powerful techniques to study very fast molecular dynamics, allowing to follow the atomic-scale motion of atoms and molecules and resolving diffusion processes on timescales from ns to sub-ps\cite{Hedgeland2009,Fouquet2010,Bahn2016,Bahn2017}. Ammonia on graphite is a fast diffusing system, accessible within the time-window of neutron time-of-flight spectroscopy. Together with van der Waals (vdW) corrected DFT calculations we show that ammonia follows a jump motion on a weakly corrugated potential energy surface.

\section{Experimental and computational details}
\subsection{Sample preparation}
\begin{figure}[htb]
\centering
\includegraphics[width = 0.48\textwidth]{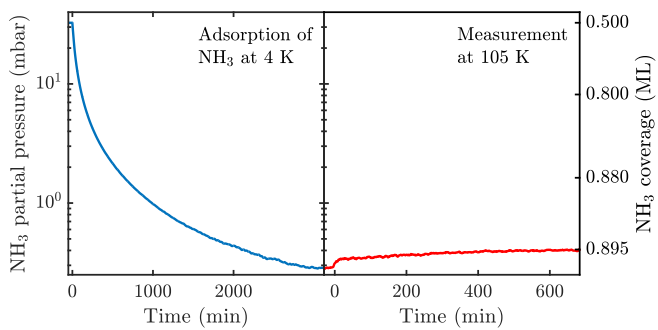}
\caption{The adsorption process of NH$_3$ on exfoliated graphite can be followed by monitoring the pressure in the connected reservoir. Left panel: Uptake during dosing from 0.5 to 0.9 ML coverage at a sample temperature of 4 K.\newline Right panel: During the measurements at 105 K desorption slowly starts to commence. However, the pressure rise corresponds to a loss of less than 1\% of the original coverage, so we can still safely assume a coverage of 0.9 ML.}
\label{fig:Uptake}
\end{figure}

We used exfoliated compressed graphite, \emph{Papyex}, a material that is widely employed for adsorption measurements due to its high specific adsorption surface area. It exhibits an effective surface area of about 25 m$^2$ g$^{-1}$ and retains a sufficiently low defect density\cite{Gilbert1998,Finkelstein2000}. In addition, exfoliated graphite samples show a preferential orientation of the basal plane surfaces. We exploited this and oriented the basal planes parallel to the scattering plane of the neutrons. We used 7.39 g of Papyex exfoliated graphite of grade N998 ($>99.8\%$ C, Carbone Lorraine, Gennevilliers, France). The prepared exfoliated graphite disks were heated to 973 K under vacuum for 4 days before transferring them into a cylindrical aluminium sample cartridge. The sample cartridge was sealed by an indium gasket and connected to a gas sorption system via a stainless steel capillary.\\
The sample temperature was controlled using a standard liquid helium cryostat. The sample was initially cooled down to 4 K and the quantity corresponding to 0.5 monolayer (ML) and 0.9 ML of ammonia gas, respectively, was dosed through the stainless steel capillary which was connected to a pressure control monitor. At monolayer coverage the area occupied by one NH$_3$ molecule corresponds to $\Sigma = 10.8~\mbox{\AA}^2$ (see \cite{Rowntree1990}). Throughout the entire experiment, connection to a 500 cm$^3$ reservoir at room temperature was maintained, for safety and monitoring purposes. In using this set-up any desorbed ammonia rises to the reservoir, where the desorbed quantity can be deduced through pressure monitoring (\autoref{fig:Uptake}).\\

\subsection{Instrumental details}
\begin{figure}[!htb]
\centering
\subfloat[Two-dimensional contour plot of the dynamic scattering function $S(Q,\Delta E)$ that was extracted from neutron TOF data obtained for exfoliated graphite covered by 0.9 ML of NH$_3$ at 94 K. The intense spot at about $Q=1.9~\mbox{\AA}^{-1}$ is due to the $(002)$ Bragg reflection from the basal plane of graphite.]{\includegraphics[width = 0.48\textwidth]{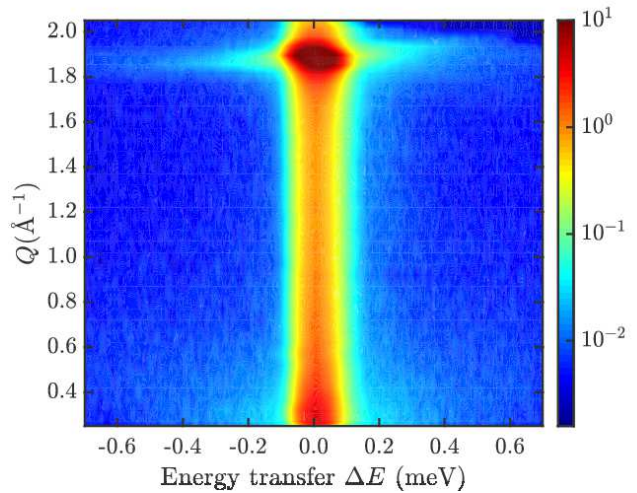}\label{fig:2DTOF}}\\
\subfloat[Comparison of the scattering functions $S(Q,\Delta E)$ at a momentum transfer of $Q=0.65~\mbox{\AA}^{-1}$ for several temperatures with the clean graphite measured at 4 K.]{\includegraphics[width = 0.42\textwidth]{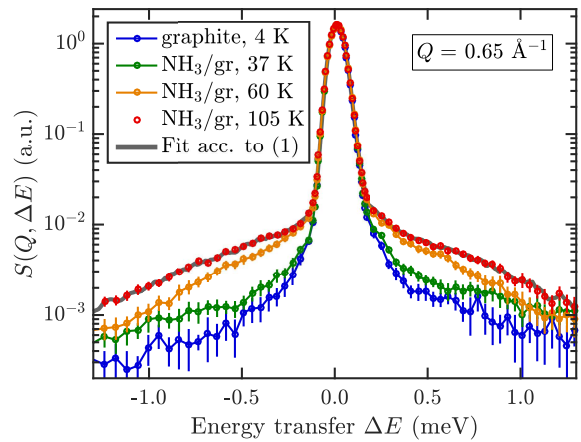}\label{fig:ExampleS}}
\caption{Neutron TOF spectra for 0.9 ML of NH$_3$ on graphite, converted to the dynamic scattering function $S(Q,\Delta E)$.}
\label{fig:TOFSQ}
\end{figure}
The measurements were performed at the IN6 time-of-flight (TOF) neutron spectrometer and the IN11 neutron spin-echo (NSE) spectrometer of the ILL\cite{doidata}. The incoming neutron wavelengths were set to $5.12~\mbox{\AA}$ and $5.5~\mbox{\AA}$, respectively, with energy resolutions at full width at
half maximum of 70 $\upmu$eV (IN6) and 1 $\upmu$eV (IN11). Neutron scattering TOF spectra of NH$_3$/graphite were obtained over a large range of temperatures: 4 K, 15 K, 25 K, 37 K, 85 K, 94 K (at 0.5 ML and 0.9 ML NH$_3$ coverages) and 105 K (only at 0.9 ML NH$_3$ coverage). Previous to the adsorption of NH$_3$, the scattering function of the graphite substrate was measured at 4 K, in order to obtain an elastic scattering resolution of the clean graphite sample.\\
The TOF spectra were converted to scattering functions, $S(Q,\Delta E)$, where $Q = | \mathbf{Q} | =  |\mathbf{k_f} - \mathbf{k_i} | $ is the momentum transfer and $\Delta E = E_f - E_i$ is the energy transfer. \autoref{fig:2DTOF} shows a two-dimensional contour plot of the dynamic scattering function $S(Q,\Delta E)$ for 0.9 ML of NH$_3$ at a temperature of 94 K. The spectrum shows an intense elastic scattering region around $\Delta E = 0$ meV which is mainly due to scattering from the graphite substrate. The broader feature surrounding the elastic band is the quasi-elastic broadening which appears due to scattering from the diffusing ammonia adsorbates.\\
A cut of the scattering function $S(Q,\Delta E)$ at $Q=0.65~\mbox{\AA}^{-1}$ is displayed in \autoref{fig:ExampleS} for several temperatures. \autoref{fig:ExampleS} shows that the quasi-elastic broadening increases with sample temperature. Up to a sample temperature of 37 K the broadening is relatively small and it is not possible to extract the quasi-elastic broadening with a reliable fit of the measured data. However, in the temperature range from 60 K to 105 K we observe a clearly discernible quasi-elastic broadening which will be used in the following to extract information about the diffusion of ammonia on exfoliated graphite.

\subsection{Computational Details}
The DFT calculations were performed using CASTEP\cite{Clark2005}, a plane wave periodic boundary condition code. The Perdew Burke Ernzerhof \cite{Perdew1996} exchange-correlation functional, with the dispersion force corrections developed by Tkatchenko and Scheffler (TS method)\cite{Tkatchenko2009}, was employed for the calculations presented in this work. The plane wave basis set was truncated to a kinetic energy cutoff of 360 eV. We have used $(4\times 4)$ and $(2\times 2)$ graphene unit cells composed of a three-layer graphene sheet to model the adsobate system at two coverages. A vacuum spacing of 20 \AA\ was imposed above the graphite surface in order to avoid interactions with the periodically repeated supercells. The substrate is frozen during the calculation and the Brillouin zone of the two unit cells are sampled with regular $(4\times 4 \times 1)$ and $(8\times 8 \times 1)$ $k$-point Monkhorst-Pack grids. The electron energy was converged up to a tolerance of $10^{-8}$ eV while the force tolerance for structural optimizations was set to 0.05 eV$/$\AA.

\section{Results and discussion}
\label{sec:TOFResults}
\begin{figure}[!thb]
     \centering
   \includegraphics[width = 0.46\textwidth]{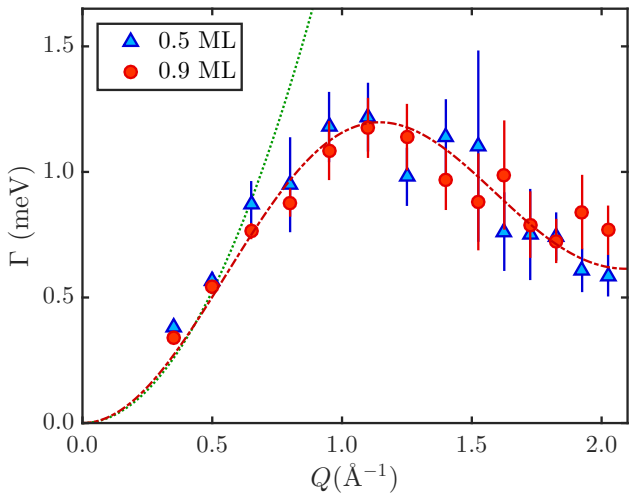}
     \caption{Extracted quasi-elastic broadening $\Gamma (Q)$ for 0.5 and 0.9 ML NH$_3$ at 94 K versus momentum transfer $Q$. The momentum transfer dependence can be described by the 2D isotropic Chudley Elliot model, Eq. \eqref{eq:CE2D}, with $l=1.4\cdot a_{gr}$ (red dash dotted curve). $\Gamma (Q)$ shows hardly any change with coverage apart from a slightly reduced broadening at small $Q$ with increasing coverage. The green dotted line shows the theoretical $\Gamma (Q)$ for Brownian motion.}
     \label{fig:QENS_Q}
\end{figure}

\begin{figure}[!thb]
     \centering
      \subfloat[Temperature dependence of the quasi-elastic broadening $\Gamma (Q)$ at 0.9 ML coverage. While the speed of the diffusion changes with temperature, the overall dependence upon $Q$ remains constant.]{\includegraphics[width = 0.46\textwidth]{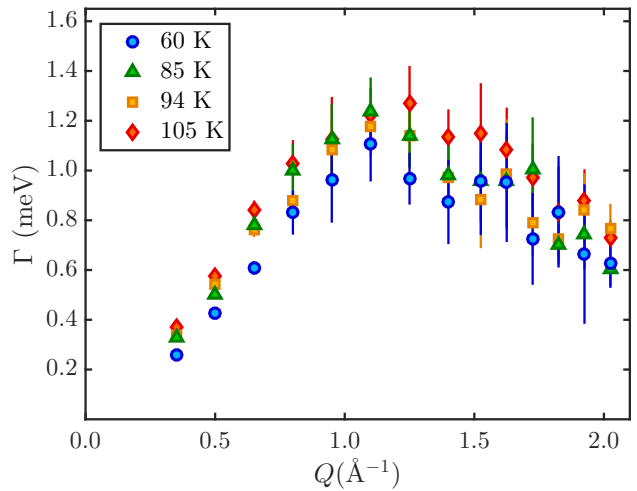}\label{fig:FitParam_09ML_allT}}\\
      \subfloat[Arrhenius plot showing the temperature dependence of the broadening $\Gamma$ at small $Q$. The activation energy for diffusion, $E_a$, is extracted from the slope of the linear fit.]{\includegraphics[width = 0.46\textwidth]{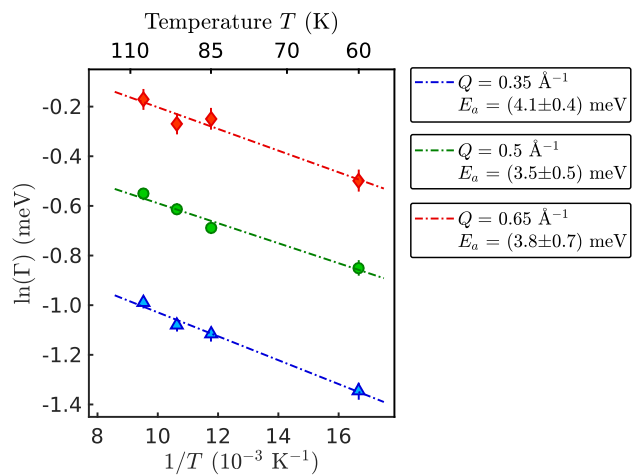}\label{fig:Arrhenius}}
     \caption{Temperature dependence of the quasi-elastic broadening for 0.9 ML of NH$_3$ on graphite.}
     \label{fig:QENS_temp}
\end{figure}

The experimentally measured scattering function $S(Q,\Delta E)$ was fitted using a convolution of the resolution function of the neutron TOF spectrometer $S_{res} (Q,\Delta E)$ with an elastic term $I_{el}(Q) \delta (\Delta E )$ and the quasi-elastic contribution $S_{inc} (Q,\Delta E)$:
\begin{widetext}
\begin{equation}
\begin{aligned}
S(Q,\Delta E) &= S_{res}(Q, \Delta E) \otimes \left[ I_{el} (Q)\delta (\Delta E ) + S_{inc} (Q,\Delta E) \right] \\
	  &= S_{res}(Q, \Delta E) \otimes \left[ I_{el} (Q)\delta (\Delta E ) + A(Q)\frac{1}{2\pi}\frac{\Gamma(Q)}{[\Gamma(Q)]^2+\Delta E^2} \right].
\label{eq:scattering_function}
\end{aligned}
\end{equation}
\end{widetext}
Here, $\delta$ represents the Dirac delta  and the quasi-elastic broadening is modelled by a Lorentzian function, where $I_{el}(Q)$ is the intensity of the elastic scattering and $A(Q)$ is the intensity of the quasi-elastic scattering. $\Gamma (Q)$ is the half width at half maximum (HWHM) of the Lorentzian. We write $S_{inc} (Q,\Delta E)$ because the quasi-elastic part of the scattering function is nearly identical to the incoherent scattering function since the coherent scattering of the graphite substrate in the considered $Q$ range is weak and the scattering of the ammonia is strongly dominated by the H atoms\cite{Bahn2016,Calvo2014}. An exemplary fit is illustrated by the thick grey line in \autoref{fig:ExampleS}.\\
The hereby extracted quasi-elastic broadening $\Gamma (Q)$ at a temperature of 94 K is plotted versus the momentum transfer $Q$ in \autoref{fig:QENS_Q}. The error bars in \autoref{fig:QENS_Q} represent the confidence intervals of the least squares fits. The error bars at small momentum transfers are invisible in the plot, since they are smaller than the size of the symbols, but for momentum transfers $Q > 0.6~\mbox{\AA}^{-1}$ they grow rapidly as $\Gamma$ approaches the widths of the spectroscopic window of the spectrometer.\\
For the case that the diffusion of the adsorbate is governed by the interaction of the molecule with a corrugated surface, its motion can be well described by the Chudley-Elliott (CE) model of jump diffusion\cite{Chudley1961,Bee1988}. The CE model assumes that a particle rests for a time $\tau$ at an adsorption site, before it moves instantaneously to another adsorption site. In the simplest case, this motion happens on a Bravais lattice and the HWHM $\Gamma (Q)$ can be expressed as:
\begin{equation}
\Gamma ( Q ) = \frac{\hbar}{N \tau} \sum_{n=1}^N \left[ 1 - \mathrm{e}^{-\mathbbm{i} \mathbf{Q}\cdot \mathbf{l}_n } \right],
\label{eq:CEmodel}
\end{equation}
where $\mathbf{l}_n$ are the corresponding jump vectors. In the case of scattering from a polycrystalline sample, isotropic angular averaging has to be
performed since the scattered neutron signal ``sees" the jumping adsorbate from all possible directions. In the case of 2D isotropy, integration in the scattering plane (over the azimuth $\varphi$) yields:
\begin{equation}
\Gamma ( Q ) = \frac{\hbar}{ \tau} \left[ 1 - J_0 \left(  Q \cdot l \cdot \sin \theta \right)  \right],
\label{eq:CE2D}
\end{equation}
where $J_0 (  Q \cdot l \cdot \sin \theta )$ is the zeroth order cylindrical Bessel function and $l$ is the average jump length. $Q \cdot \sin \theta$ is the component of the scattering vector in the plane of diffusion, and $\theta$ the angle between $\mathbf{Q}$ and the normal to this plane\cite{Lechner1995}. Papyex consists of planes with an inclination that is normally distributed around $\theta = 90^{\circ}$ with a HWHM of about $15^{\circ}$\cite{Gilbert1998}. This has been taken into account by numerical integration of \eqref{eq:CE2D}.\\
It should be noted that the isotropic averaging is only an approximation and it omits the fact that for a correct isotropic averaging one needs to integrate over the $S(Q,\Delta E)$ rather than the broadening $\Gamma(Q)$, which produces in general a non-Lorentzian QENS broadening\cite{Jobic2000,Bahn2016}. However, the deviation from the Lorentz distribution is mainly caused due to scattering processes which occur almost perpendicular to the plane of diffusion. While this contribution should not be neglected in the case of three-dimensional polycrystalline materials, in the case of Papyex the scattering vector $\mathbf{Q}$ is approximately parallel to the $(0001)$ basal plane of graphite as mentioned above. Hence we will rely on the approximate solution \eqref{eq:CE2D}, which produces very good results.\\
\eqref{eq:CE2D} is then fitted to the experimentally determined broadening $\Gamma (Q)$ using an iterative generalized least squares algorithm with weights (and a numerical integration over $\theta$). The red dash-dotted line in \autoref{fig:QENS_Q} shows that \eqref{eq:CE2D} fits the data very well for $l=(3.45\pm 0.02)~\mbox{\AA}$ and $\tau = (0.85\pm 0.08)~\mbox{ps}$. From the momentum transfer dependence we can clearly exclude other types of motion. E.g. ballistic diffusion, which represents a two dimensional ideal gas, is characterised by a linear dependence of $\Gamma(Q)$. Moreover, Brownian diffusion which describes a continuous motion, is characterised by a square law dependence of the momentum transfer (green dotted line in \autoref{fig:QENS_Q}) and cannot reproduce the momentum transfer dependence of the broadening.\\
Note that the average jump distance ($l=3.45~\mbox{\AA}$) corresponds to $1.4 \, a_{gr}$ where $a_{gr}$ is the graphite lattice constant. Hence the average jump length suggests that a significant number of long jumps occurs at this temperature. Using the residence time $\tau$ and the average jump length $l$ Einstein's equation for diffusion (in the two-dimensional case) can be used to determine the diffusion constant $D$\cite{Bee1988}:
\begin{equation}
D = \frac{ \left\langle  l \right\rangle^2 }{4 \tau}
\label{eq:EinsteinDiff}
\end{equation}
with the mean jump length $\langle l \rangle$. Using \eqref{eq:EinsteinDiff} we obtain a diffusion constant of $D=(3.9 \pm 0.4) \cdot 10^{-8}~\mbox{m}^2 /\mbox{s}$ at 94 K. The diffusion constant for ammonia adsorbed on graphitized carbon black has been determined to range from $D=0.6 \cdot 10^{-8}~\mbox{m}^2 /\mbox{s}$ at 180 K to $D=6 \cdot 10^{-8}~\mbox{m}^2 /\mbox{s}$ at 230 K using NMR\cite{Tabony1979} with similar values at 205 K using neutron scattering\cite{Gamlen1979}. Considering that these values were determined at much higher temperatures (where ammonia on graphite will already have been completely desorbed) and for a different substrate, the diffusion constants are within the same order of magnitude compared to our results.\\
The diffusion of small molecules on graphite and graphene has been mainly treated by theoretical approaches where typically a fast diffusion process is predicted\cite{Ma2011,Tang2011}. E.g. Ma \emph{et al.}\cite{Ma2011} report that H$_2$O adsorbed on graphene undergoes an ultra-fast diffusion process at 100 K with $D=6 \cdot 10^{-9}~\mbox{m}^2 /\mbox{s}$. The value determined for ammonia in our study is even one order of magnitude larger showing that the diffusion of ammonia on graphite is a very rapid process. Compared to other experimental studies it is about the same size compared to the jump diffusion of molecular hydrogen (H$_2$) on graphite\cite{Haas2009,Bahn2016} and again one order of magnitude larger than the diffusion constant found for benzene (C$_6$H$_6$) on graphite\cite{Hedgeland2009}.\\
As a next step we consider the coverage and temperature dependence of the diffusion process. Unfortunately, the signal-to-noise ratio and the difference between the scattering function and the resolution function is too small for the data measured at 0.5 ML coverage to extract a reliable quasi-elastic broadening. The only exception is the highest temperature (94 K), measured at this coverage. This is due to the fact that with increasing temperature the broadening becomes larger, as one would expect for an activated motion. \autoref{fig:QENS_Q} shows a comparison of the quasi-elastic broadening $\Gamma (Q)$ for 0.5 and 0.9 ML of NH$_3$ coverage as a function of momentum transfer $Q$. One may anticipate a slightly reduced broadening at the higher coverage and thus a smaller hopping rate, which is however, only discernible at small $Q$ due to the uncertainties. In general the experiments show no significant coverage dependence within the experimental uncertainties. Hence we cannot quantify the respective contributions of the molecule-molecule collisions or the molecule-surface interactions to the diffusion process.\\ 
In \autoref{fig:FitParam_09ML_allT} the quasi-elastic broadening $\Gamma (Q)$ is plotted for all temperatures measured at an NH$_3$ coverage of 0.9 ML. The broadening and hence the hopping rate becomes larger with increasing temperature, but the overall dependence upon $Q$, i.e., the hopping distance, remains largely constant.\\
While at high $Q$ the uncertainties in \autoref{fig:FitParam_09ML_allT} are too large to extract a meaningful temperature dependence, we can use the temperature dependence of $\Gamma$ at small $Q$, i.e., for long range diffusion, to obtain a diffusion barrier. For a thermally activated processes, Arrhenius' law predicts a temperature dependence of the broadening $\Gamma$, as:
\begin{equation}
\Gamma = \Gamma_0 \, \text{e}^{-\frac{E_a}{k_B \, T}},
\label{eq:Arrhenius}
\end{equation}
where $\Gamma_0$ is the pre-exponential factor, $E_a$ is the activation energy for diffusion, $k_B$ the Boltzmann constant and $T$ the sample temperature. Taking the natural logarithm of \eqref{eq:Arrhenius} results in a linear relationship between the inverse of the temperature, $1/T$, and the natural logarithm of the broadening $\Gamma$.\\
\autoref{fig:Arrhenius} shows such an Arrhenius plot of the broadening $\Gamma$ for the three lowest momentum transfers $Q$. The activation energy, extracted form the linear fit varies between 3.5 and 4.1 meV giving rise to a mean value of $E_a = (3.8 \pm 0.7)~\mbox{meV}$.\\
Note that the hereby determined diffusion barrier is smaller than the thermal energy ($k_B T$) of the substrate, while on the other hand the thermal energy is still significantly below the desorption energy. Other experimental examples for the occurrence of jump-like diffusion in the case of a very low potential energy barrier include the case of Cs on Cu(001)\cite{Jardine2007}. Nevertheless, it is quite unusual to observe hopping motion for a system with such a weakly corrugated potential energy surface. It suggests that substantial energy dissipation channels must be present in the ammonia/graphite system (e.g., by molecular collisions or by energy dissipation to the surface), in contrast to the diffusion
of flat hydrocarbons such as pyrene on graphite\cite{Calvo2016}.\\
In general, at temperatures higher than the diffusion barrier height, the time spent by the adsorbate near the minimum of the adsorption potential is comparable to the time in the in-between regions. In this case both diffusive and vibrational motions, associated with a temporary trapping of an adsorbate inside the surface potential well, contribute to the quasielastic broadening and are coupled\cite{Chen1993}. As theoretically proposed by Mart\'{i}nez-Casado \emph{et al.}\cite{MartinezCasado2007} in a generalised model for the quasi-elastic broadening, a combination of both cases should give rise to a more complicated dependence of the broadening on the momentum transfer due to the diffusive hopping motion and the friction parameter $\eta$. As shown by Jardine \emph{et al.}\cite{Jardine2004}, friction may become more apparent in the broadening due to these vibrational motions, whereas the contribution of the effect to energy dissipation during diffusion cannot be decoupled due to the final energy resolution of the instrument. The internal degrees of freedom of the adsorbed molecule may even further complicate the underlying microscopic processes\cite{Lechner2013}.\\
However, based on the approach by Mart\'{i}nez-Casado \emph{et al.}\cite{MartinezCasado2007}, we can use the fact that the CE model contains Brownian diffusion as a long range diffusion limit, to obtain a crude estimate for the friction. For $Q\rightarrow 0$ the broadening converges to a parabola, i.e. the broadening approaches the same momentum transfer dependence as for Brownian motion, where the atomic-scale friction $\eta$ can be directly extracted using Einstein's relation\cite{Fouquet2010} as used in the fluctuation-dissipation theorem by Kubo\cite{Fouquet2010,Calvo2014}:
\begin{equation}
D = \frac{k_B T}{\eta m},
\label{eq:EinsteinDiff2}
\end{equation}
where $m$ is the mass of the ammonia molecule. Using this approximation we obtain an estimate of the atomic-scale friction of $\eta = 1.2~\mbox{ps}^{-1}$ from the data in \autoref{fig:QENS_Q}, which is a medium value for the atomic-scale friction compared to previous studies\cite{Jardine2004,Jardine2007,Lechner2013}.\\
We would like to stress that the result should be taken with care and can only serve as a crude estimate. Friction in surface diffusion processes can be caused by a variety of energy dissipation channels, including also interactions between the adsorbates and interaction with the substrate. Since the measurements were performed close to the monolayer regime, the friction parameter extracted from the fitting of the quasi-elastic broadening to a parabola at low momentum transfers cannot be written as a simple sum of contributions to the energy dissipation\cite{MartinezCasado2010}. It is rather an averaged friction parameter which is related to the energy dissipation frequency of a single molecule diffusing on the basal plane of graphite and interacting with the surface phonon bath and its neighbouring molecules.\\
Nonetheless it suggests that friction plays a significant role in the NH$_3$/graphite system. Indeed, for a system with non-negligible friction, one would expect that for each single jump an energy equivalent to the height of the barrier is dissipated\cite{Boisvert1996,Hershkovitz1999}. I.e. energy dissipation via frictional coupling is likely to be responsible for the occurrence of the hopping motion. On the other hand with increasing thermal energy compared to the potential energy surface, more and more long jumps start to set in during jump diffusion\cite{Linderoth1997,Hershkovitz1999,Ferron2009,MiretArtes2005}, which is evident from the experimental data, since the best fit Chudley-Elliott model gives an average jump length of $1.4\, a_{gr}$.\\
Note that a similar diffusive motion was observed for molecular hydrogen on graphite with jump diffusion and also a very low activation energy\cite{Haas2009,Bahn2016,Petucci2013}. Although the role of atomic-scale friction was not explicitly discussed in those cases, it suggests together with the results presented in our study, that friction may be partly caused by the geometry of the molecule when compared to the flat-lying polycyclic aromatic hydrocarbons which closely resemble the structure of the graphite substrate\cite{Calvo2016,Hedgeland2009,Guo2015}.\\
Finally, the occurrence of long jumps makes the determination of a meaningful activation energy challenging since under these circumstances jumps start to become correlated as shown in theoretical studies\cite{Ferron2004,Boisvert1996}. In the case of exfoliated graphite this is further complicated by the azimuthal averaging as described above. Nevertheless we will use this value as a rough estimate for the diffusion of ammonia on graphite and attempt in the following to compare our experimental results with DFT calculations. 

\subsection{DFT results}
\begin{figure}[!thb]
\centering
\includegraphics[width = 0.42\textwidth]{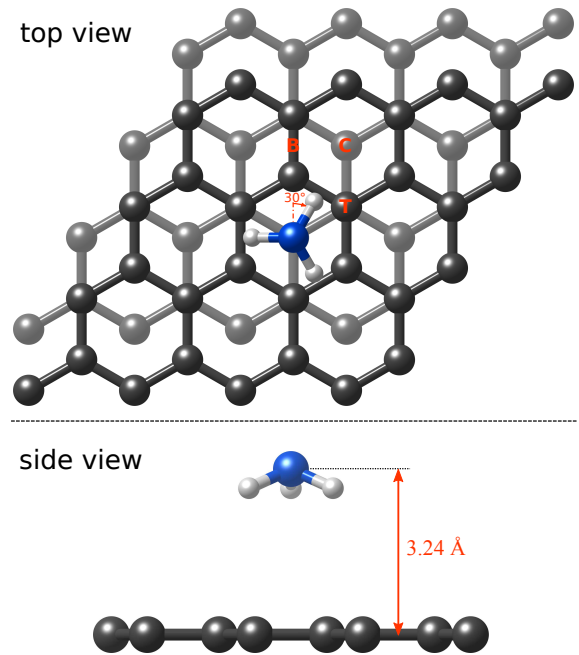}
\caption{Geometry of the NH$_3$/graphite system investigated in this study. The high symmetry adsorption positions with respect to the graphite lattice are labelled as T: on-top; B: onbond or bridge and C: centre. The most favourable adsorption site according to vdW corrected DFT is for NH$_3$ at the centre position with the rotation axis perpendicular to the surface and the
hydrogen atoms directed towards the onbond sites.}
\label{fig:AmmoniaDFT}
\end{figure}

\begin{figure}[!thb]
\centering
\subfloat[Downwards configuration]{\includegraphics[width = 0.24\textwidth]{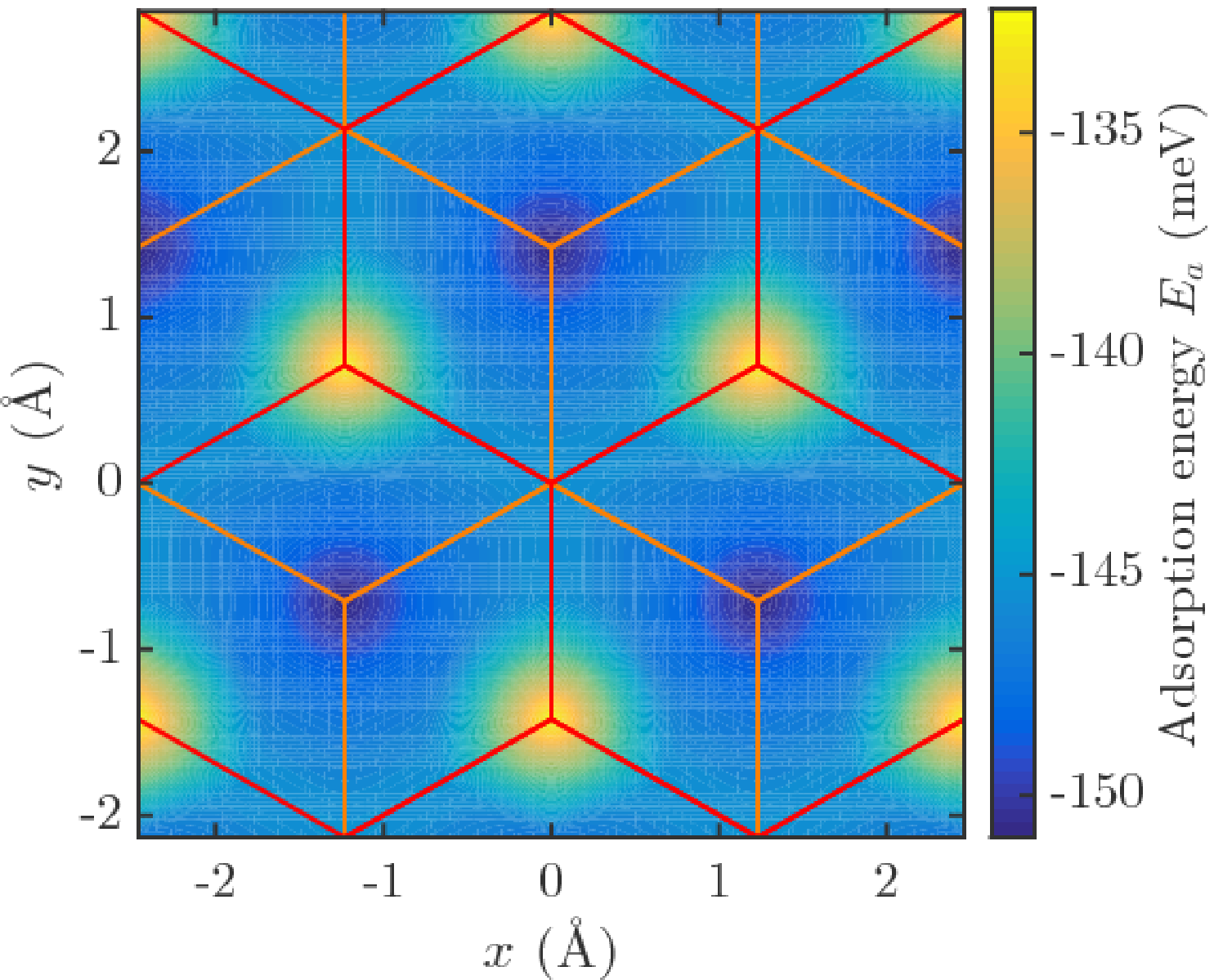}\label{fig:DFT_PES1}}\hfill
\subfloat[Upwards configuration]{\includegraphics[width = 0.24\textwidth]{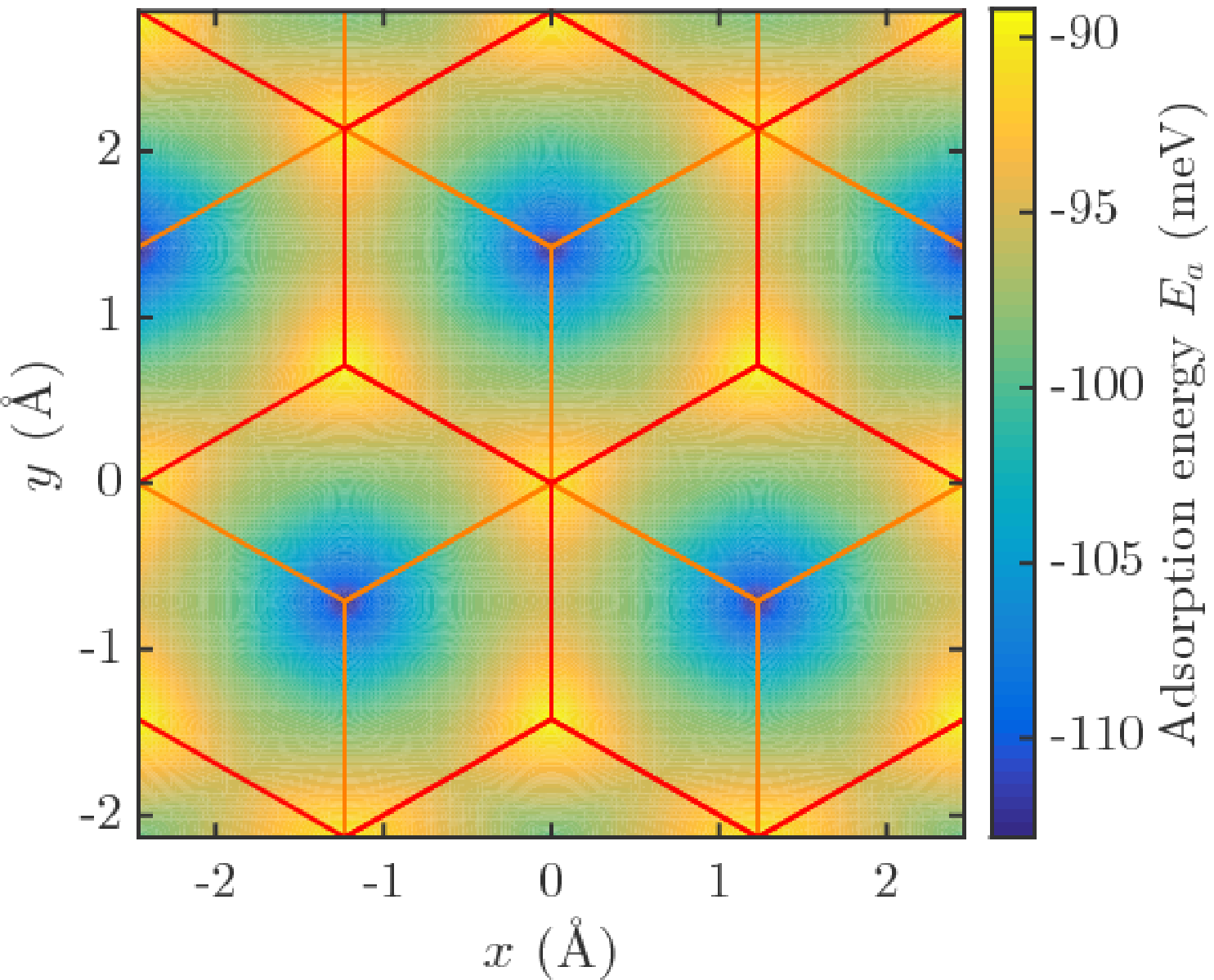}\label{fig:DFT_PES2}}
\caption{Comparison of the potential energy surface as obtained by the vdW corrected DFT for NH$_3$ in the downwards and upwards configuration. Both calculations are for the minimum energy  rotation of $30^{\circ}$ and at a coverage of approximately 1 ML NH$_3$. The red and orange lines represent the first and second layer of the graphite substrate, respectively.}
\label{fig:DFT_PES}
\end{figure}

We have studied the adsorption of NH$_3$ on graphite for a large number of different adsorption geometries. Those include 6 different adsorption sites within the graphite unit cell, the orientation of the molecule with the hydrogen atoms pointing upwards (U) or downwards (D) as well as three different rotations around the axis perpendicular to the surface. \autoref{fig:AmmoniaDFT} shows the energetically most favourable adsorption site, with the molecule located at the C site (centre) and the H-atoms pointing towards the surface, directed towards the onbond sites.\\ 
Based on the vdW corrected DFT calculations the adsorption energy of a single NH$_3$ molecule on graphite is 173 meV, which is slightly reduced to 151 meV in the high coverage regime (about 1 ML). Note that the adsorption energy is much closer to the experimentally found values from TDS than in previous DFT calculations which yielded adsorption energies in the order of 25-30\% of the experimentally determined value. Hence it shows the importance of vdW interactions in this system and that previous DFT results (without vdW interactions) should be taken cautiously when trying to make predictions.\\
Interestingly B\"{o}ttcher \emph{et al.}\cite{Boettcher2011} obtain a similar adsorption energy of 146 meV for NH$_3$ on graphene/Ni(111) from vdW corrected DFT, however, the molecule is adsorbed in the upwards configuration on graphene/Ni(111). On the other hand, recent X-ray absorption spectroscopy measurements provided evidence for a chemical contribution to the adsorption bond in the case of NH$_3$ adsorbed on graphene/Ni(111)\cite{Boettcher2017}. Hence it is possible that due to the present metal substrate the adsorption geometry of the ammonia molecule on graphene/Ni(111) changes compared to ammonia adsorbed on graphite.\\
\begin{table}[htb]
\caption{The adsorption energy $E_a$ and the energy difference $\Delta E_a$ relative to the most favourable adsorption site for NH$_3$ on graphite. The six different adsorption geometries are with the H-atoms pointing upwards (U) or downwards (D) and the centre (C), top (T) and bridge (B) adsorption site.}
\centering
\begin{tabular}{c c . .}
\hline
Orientation & Position & \multicolumn{1}{l}{ $E_a$ (eV) } & \multicolumn{1}{l}{ $\Delta E_a$ (meV) } \\ 
\hline
D & T & -0.144 & 7 \\ 
D & B & -0.145 & 6 \\ 
D & C & -0.151 & 0 \\ 
U & T & -0.089 & 62 \\
U & B & -0.095 & 56 \\ 
U & C & -0.113 & 38 \\ 
\hline
\end{tabular}
\label{tab:DFTAds}
\end{table}
\autoref{tab:DFTAds} summarises six arrangements where the molecule is placed in the high symmetry positions (T, B, and C) at a rotation of $30^{\circ}$ for an ammonia coverage of about 1 ML. For the complete set (including all considered adsorption geometries and coverages) please refer to the supplementary information. We conclude from \autoref{tab:DFTAds} that the downwards configuration is definitively favoured with respect to the upwards configuration, regardless of the adsorption site. For the down configuration the energy differences between different adsorption sites are in general extremely small. Moreover, the distance of the molecule with respect to the surface does not vary significantly, e.g., for a given rotation angle and downwards orientation the minimum distance is 3.24 \AA\ at the C site and the maximum is 3.26 \AA\ at the B site.\\
Hence, the DFT calculations confirm that the diffusion of ammonia on graphite should be governed by a weakly corrugated potential energy surface. It can also be seen from \autoref{fig:DFT_PES} which shows a contour plot of the potential energy surface for NH$_3$ adsorbed on different positions of the graphite substrate. The adsorption energies for both the upwards and the downwards configuration are illustrated, as extracted from the vdW corrected DFT calculations with the minimum energy rotation of $30^{\circ}$ and at a coverage of approximately 1 ML NH$_3$. For the downwards configuration, \autoref{fig:DFT_PES1}, the top site located above the second layer carbon atom is energetically less favourable by a significant amount but all other adsorption positions vary only by several meV. Based on the ``static snapshots'' i.e. the energy differences between the adsorption sites from vdW corrected DFT (\autoref{tab:DFTAds} and \autoref{fig:DFT_PES1}) the diffusion barrier would be 6 meV which is in good agreement with the value extracted from the experimental data. According to this the most likely trajectory would be from the C site via the B site to the next C site.\\ 
Furthermore, we have also calculated the energy difference for nitrogen inversion (the umbrella or symmetric deformation vibration mode) on graphite. Here, the energy difference between the up and down NH$_3$ configuration in a given position can only serve as a lower limit to the ``real'' inversion barrier and gives 38 meV for 1 ML of NH$_3$ in our case. Therefore we have also calculated the transition state structure for NH$_3$ inversion on the global minimum for both the $(2\times 2)$ and $(4\times4)$ cells. At lower coverage the barrier is 157 meV (starting from the down configuration) and 142 meV (starting from the up configuration). At higher coverage, the barriers are reduced to 132 meV and 94 meV, respectively. Since the down and up configurations are not symmetrical, there is a slight difference in the barrier from the down and up structures.\\
There is quite a substantial activation energy change when going to the higher coverage. We suspect that this change may be caused by repulsive steric interactions between the hydrogen atoms of two adjacent NH$_3$ molecules. In general the barrier is in line with the values reported for other systems with adsorbed ammonia. E.g the energy of this mode is typically between 130-145 meV for NH$_3$ adsorbed on metal surfaces\cite{Parmeter1988,Pascual2003}. For NH$_3$ on HOPG the umbrella mode could only be observed in the multilayer case where the value is similar to the one for solid ammonia\cite{Bolina2005}. 
upon adsorption on graphite. 

\subsection{Spin-echo measurements}
The neutron spin-echo experiments for deuterated ammonia (ND$_3$) at a surface coverage of 0.9 ML were conducted on IN11 for sample temperatures of 2 K (resolution) and for 60 K, 85 K, 94 K and 105 K. The NSE measurement delivers the development of the space correlation function with time $t$, i.e., the normalised intermediate scattering function $S(Q,t)/S(Q,0)$\cite{Fouquet2010,Mezei1980}. This function can also be obtained by Fourier transforming the scattering function $S(Q,\Delta E$). Converting the quasi-elastic broadening determined in \autoref{sec:TOFResults} to a broadening in time gives rise to $\tau \approx 1~\mbox{ps}$ at $Q = 0.5~\mbox{\AA}^{-1}$. This is below the spectral acceptance window of IN11 and the corresponding decay does not appear in the IN11 spectra. Nevertheless, the spin-echo measurements show that there is no additional motion at longer timescales, confirming the fast diffusion process seen in the TOF measurements (see also the supplementary information).\\

\section{Summary and conclusion}
We have studied the diffusion of ammonia on exfoliated graphite using quasi-elastic neutron scattering. The dependency of the quasielastic broadening on the  momentum transfer shows that ammonia follows a hopping motion on the basal plane of graphite. The diffusion constant at 94 K was determined as $D=(3.9\pm0.4) \cdot 10^{-8}~\mbox{m}^2 /\mbox{s}$ suggesting that the diffusion of ammonia on graphite is a very rapid process, comparable to the diffusion of molecular hydrogen and much faster than the diffusion of larger molecules, such as benzene. Considering in particular the mass of the molecule, together with the unusual tilted NH$-\pi$ bonding, makes the observed diffusion in this system uniquely fast. In terms of possible applications for gas sensing purposes, it implies that after adsorption the kinetics on the surface should not be the limiting factor.\\
The activation energy extracted from the temperature dependence of the quasielastic broadening is about 4 meV. The combination of jump diffusion and a low activation energy suggests that NH$_3$/graphite is a system with a rather unusual combination of a weakly corrugated potential energy surface together with a significant friction. The combination of jump diffusion and a low activation energy suggests that NH$_3$/graphite is a system with a rather unusual combination of a weakly corrugated potential energy surface together with a significant friction. We hope that our work will initiate further theoretical investigations in order to address this interesting finding.\\
The calculated potential energy surfaces is extremely flat for a given orientation of the molecule. The configuration of the adsorbate with the reverse polarity (NH bonds pointing upwards) is energetically unfavourable, therefore breaking the symmetry of the umbrella  inversion mode. Furthermore, the adsorption energy of ammonia on graphite is determined as 173 meV from DFT, much closer to the experimental value compared to previous DFT calculations without dispersion corrections. The close agreement between the calculated adsorption energy, diffusion barrier and the experimental results confirm the accuracy of the TS dispersion corrections scheme for vdW bonded systems on graphite.

\section{Acknowledgement}
One of us (A.T.) acknowledges financial support provided by the FWF (Austrian Science Fund) within the project J3479-N20. The authors would like to thank E. Bahn for many helpful discussions. M.S. is grateful for the support from the Royal Society. This work used the ARCHER UK National Supercomputing Service via the membership of the UK's HEC Materials Chemistry Consortium which is funded by the EPSRC (EP/L000202). The authors acknowledge the generous provision of neutron beam time at the ILL. 

\bibliography{literature}

\clearpage
\onecolumngrid

\renewcommand{\thepage}{\arabic{apppage}}
\pagenumbering{arabic}

\setcounter{section}{0}
\renewcommand{\thesection}{S\arabic{section}}
\section*{\large{Supplementary Information:\\ Ultrafast Molecular Transport on Carbon Surfaces:\\ The Diffusion of Ammonia on Graphite}}

\setcounter{table}{0}
\setcounter{figure}{0}
\makeatletter
\renewcommand{\thetable}{S\arabic{table}}
\renewcommand{\thefigure}{S\arabic{figure}}

\section{Full Set of DFT calculations}
\begin{figure}[!ht]
\centering
\subfloat[Top view]{\includegraphics[width = 0.41\textwidth]{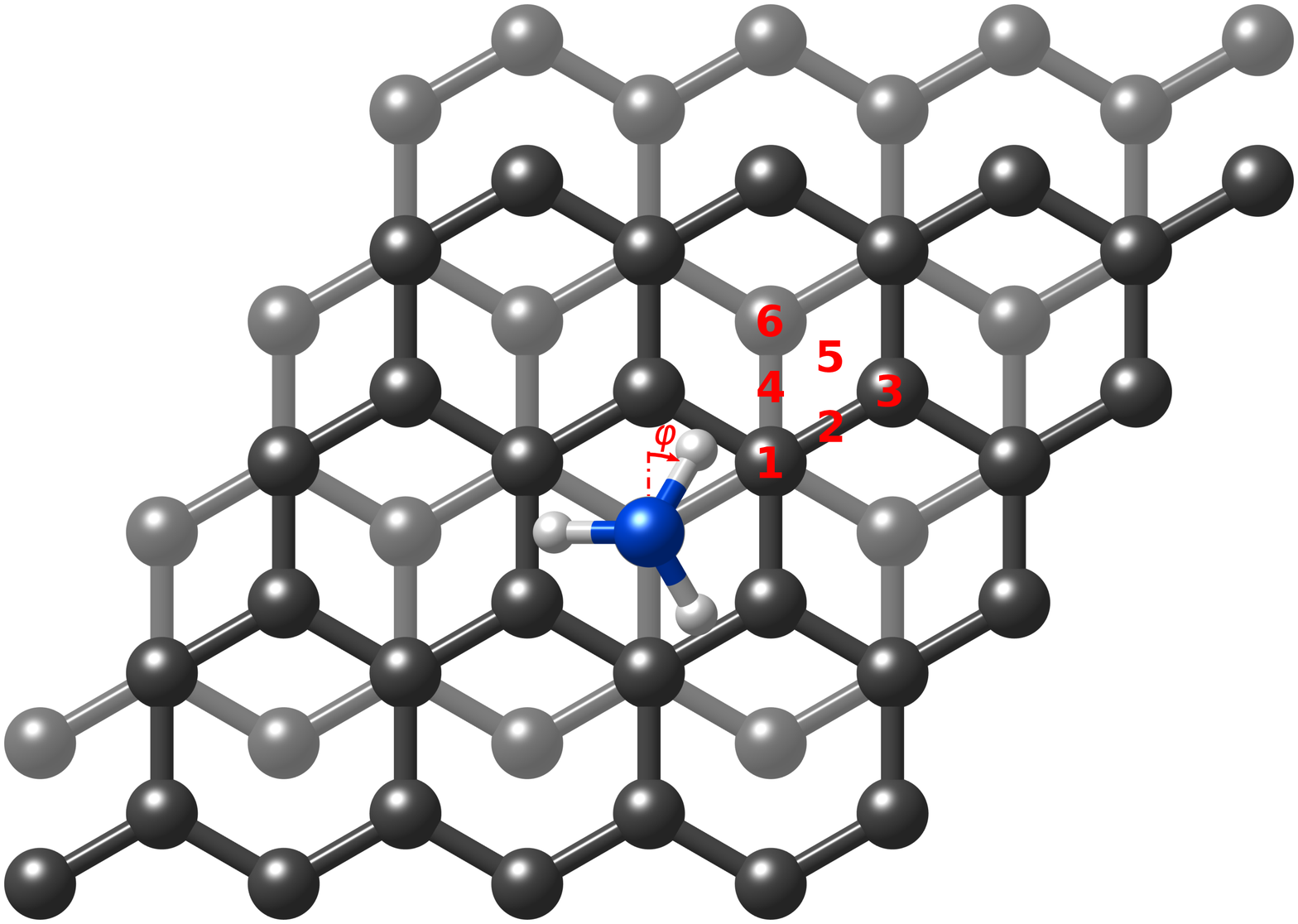}}\\[0.1cm]
\subfloat[Side view showing the nitrogen inversion]{\includegraphics[width = 0.45\textwidth]{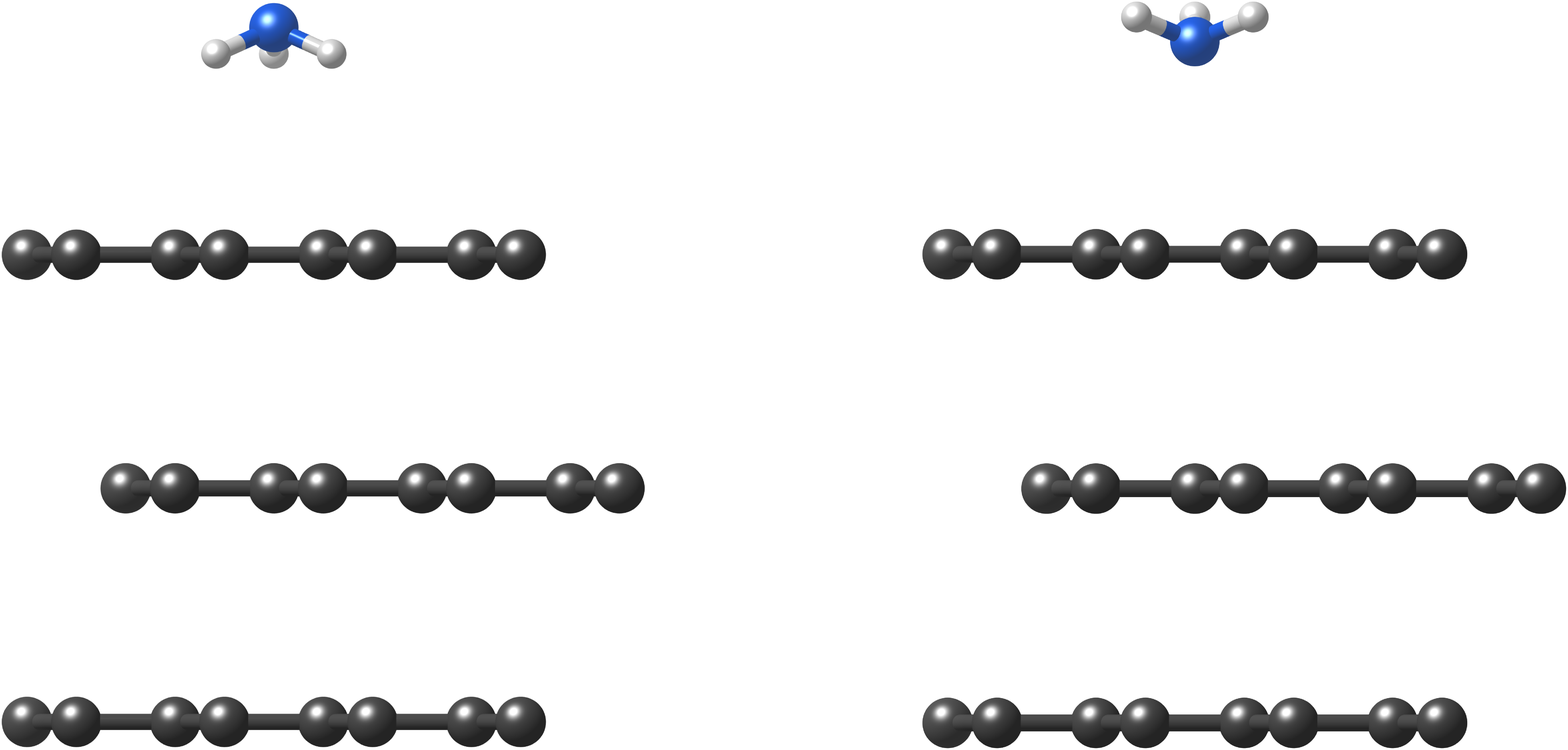}}
\caption{Geometry of the NH$_3$/graphite system investigated in this study. The considered adsorption positions with respect to the graphite lattice are labelled with the numbers $1,2,... 6$. Three different rotations around the axis perpendicular to the surface $\varphi$ have been considered as well. Finally, the adsorption of the molecule with the hydrogen atoms pointing towards the surface and in an upwards configuration were considered as well.}
\label{fig:ammoniaDFT2}
\end{figure}

In this section we present the data for all adsorption geometries / configurations of ammonia on graphite which have been calculated using van-der-Waals corrected DFT. We have calculated the adsorption energy of NH$_3$ adsorbed on several different positions on top of graphite, for several rotational angles as well as with hydrogen atoms of the molecule pointing upwards (U) or downwards (D).\\
For the 6 considered adsorption positions and the rotation of the molecule $\varphi$ see the illustration in \autoref{fig:ammoniaDFT2}. Note that the difference between the two top sites (position 1 and position 3) is given by the carbon atom sitting in the second layer underneath the top site. The results for a $(2 \times 2)$ unit cell are summarised in \autoref{tab:DFT_ammonia1} and the results for a $(4 \times 4)$ unit cell are given in \autoref{tab:DFT_ammonia2}.\\

\begin{table}[ht]
\caption{Summary of the vdW corrected DFT calculations for NH$_3$ on a $(2 \times 2)$ graphite unit cell.\\
Orient ... Orientation of NH$_3$ with the H-atoms pointing downwards (D) or upwards (U)\\
Rot ... Rotation of NH$_3$ along the $z$-axis\\
Pos ... Adsorption position on graphite (\autoref{fig:ammoniaDFT2})\\
$E_a$ ... Calculated adsorption energy \\
$\Delta E$ ... Energy difference with respect to the most favourable adsorption geometry\\
$\Delta E_{inv}$ ... Energy difference for nitrogen inversion}
\centering
\begin{tabular}{c . . . . . .}
\hline
Orient & \multicolumn{1}{c}{Rot. $(^{\circ})$} & \multicolumn{1}{c}{Pos.} & \multicolumn{1}{c}{$E_a$ (eV)} & \multicolumn{1}{c}{$\Delta E$ (meV)} & \multicolumn{1}{c}{$\Delta E_{inv}$ (meV)} \\ 
\hline
D & 0 & 1 & -0.140 & 11 & 57 \\
D & 0 & 2 & -0.138 & 14 & 55 \\ 
D & 0 & 3 & -0.132 & 19 & 55 \\ 
D & 0 & 4 & -0.140 & 11 & 49 \\ 
D & 0 & 5 & -0.139 & 12 & 49 \\ 
D & 0 & 6 & -0.144 & 8 & 39 \\ 
D & 30 & 1 & -0.145 & 7 & 55 \\ 
D & 30 & 2 & -0.145 & 6 & 50 \\ 
D & 30 & 3 & -0.132 & 19 & 43 \\ 
D & 30 & 4 & -0.147 & 4 & 48 \\ 
D & 30 & 5 & -0.147 & 4 & 48 \\ 
D & 30 & 6 & -0.151 & 0 & 38 \\ 
D & 60 & 1 & -0.134 & 17 & 55 \\ 
D & 60 & 2 & -0.137 & 14 & 54 \\ 
D & 60 & 3 & -0.138 & 13 & 52 \\ 
D & 60 & 4 & -0.138 & 13 & 48 \\ 
D & 60 & 5 & -0.140 & 11 & 49 \\ 
D & 60 & 6 & -0.143 & 9 & 38 \\ 
U & 0 & 1 & -0.083 & 68 &  \\ 
U & 0 & 2 & -0.083 & 68 &  \\ 
U & 0 & 3 & -0.078 & 73 &  \\ 
U & 0 & 4 & -0.091 & 60 &  \\ 
U & 0 & 5 & -0.090 & 61 &  \\ 
U & 0 & 6 & -0.105 & 46 &  \\ 
U & 30 & 1 & -0.090 & 62 &  \\ 
U & 30 & 2 & -0.095 & 56 &  \\ 
U & 30 & 3 & -0.089 & 62 &  \\ 
U & 30 & 4 & -0.099 & 52 &  \\ 
U & 30 & 5 & -0.099 & 52 &  \\ 
U & 30 & 6 & -0.113 & 38 &  \\ 
U & 60 & 1 & -0.080 & 72 &  \\ 
U & 60 & 2 & -0.083 & 68 &  \\ 
U & 60 & 3 & -0.086 & 65 &  \\ 
U & 60 & 4 & -0.090 & 61 &  \\ 
U & 60 & 5 & -0.092 & 60 &  \\ 
U & 60 & 6 & -0.105 & 47 &  \\ 
\hline
\end{tabular}
\label{tab:DFT_ammonia1}
\end{table}

\begin{table}[ht]
\caption{Summary of the vdW corrected DFT calculations for NH$_3$ on a $(4 \times 4)$ graphite unit cell.\\
Orient ... Orientation of NH$_3$ with the H-atoms pointing downwards (D) or upwards (U)\\
Rot ... Rotation of NH$_3$ along the $z$-axis\\
Pos ... Adsorption position on graphite (\autoref{fig:ammoniaDFT2})\\
$E_a$ ... Calculated adsorption energy\\
$\Delta E$ ... Energy difference with respect to the most favourable adsorption geometry\\
$\Delta E_{inv}$ ... Energy difference for nitrogen inversion}
\centering
\begin{tabular}{c . . . . . .}
\hline
Orient & \multicolumn{1}{c}{Rot. $(^{\circ})$} & \multicolumn{1}{c}{Pos.} & \multicolumn{1}{c}{$E_a$ (eV)} & \multicolumn{1}{c}{$\Delta E$ (meV)} & \multicolumn{1}{c}{$\Delta E_{inv}$ (meV)} \\ 
\hline
D & 0 & 1 & -0.164 & 10 & 34 \\ 
D & 0 & 2 & -0.161 & 12 & 31 \\ 
D & 0 & 3 & -0.167 & 6 & 42 \\ 
D & 0 & 4 & -0.165 & 9 & 26 \\ 
D & 0 & 5 & -0.162 & 11 & 25 \\ 
D & 0 & 6 & -0.170 & 3 & 16 \\ 
D & 30 & 1 & -0.163 & 10 & 31 \\ 
D & 30 & 2 & -0.165 & 9 & 30 \\ 
D & 30 & 3 & -0.163 & 10 & 31 \\ 
D & 30 & 4 & -0.167 & 6 & 25 \\ 
D & 30 & 5 & -0.167 & 6 & 25 \\ 
D & 30 & 6 & -0.173 & 0 & 15 \\ 
D & 60 & 1 & -0.168 & 5 & 41 \\ 
D & 60 & 2 & -0.161 & 12 & 31 \\ 
D & 60 & 3 & -0.163 & 10 & 34 \\ 
D & 60 & 4 & -0.162 & 11 & 25 \\ 
D & 60 & 5 & -0.165 & 9 & 26 \\ 
D & 60 & 6 & -0.171 & 2 & 17 \\ 
U & 0 & 1 & -0.130 & 44 &  \\ 
U & 0 & 2 & -0.130 & 43 &  \\ 
U & 0 & 3 & -0.126 & 48 &  \\ 
U & 0 & 4 & -0.139 & 34 &  \\ 
U & 0 & 5 & -0.137 & 36 &  \\ 
U & 0 & 6 & -0.155 & 18 &  \\ 
U & 30 & 1 & -0.132 & 41 &  \\ 
U & 30 & 2 & -0.135 & 39 &  \\ 
U & 30 & 3 & -0.132 & 42 &  \\ 
U & 30 & 4 & -0.142 & 31 &  \\ 
U & 30 & 5 & -0.142 & 31 &  \\ 
U & 30 & 6 & -0.158 & 15 &  \\ 
U & 60 & 1 & -0.127 & 47 &  \\ 
U & 60 & 2 & -0.130 & 43 &  \\ 
U & 60 & 3 & -0.129 & 44 &  \\ 
U & 60 & 4 & -0.138 & 36 &  \\ 
U & 60 & 5 & -0.139 & 35 &  \\ 
U & 60 & 6 & -0.154 & 19 &  \\
\hline
\end{tabular}
\label{tab:DFT_ammonia2}
\end{table}

\clearpage

\section{Neutron spin-echo measurements}
\begin{figure}[htb]
     \centering
     \includegraphics[width = 0.48\textwidth]{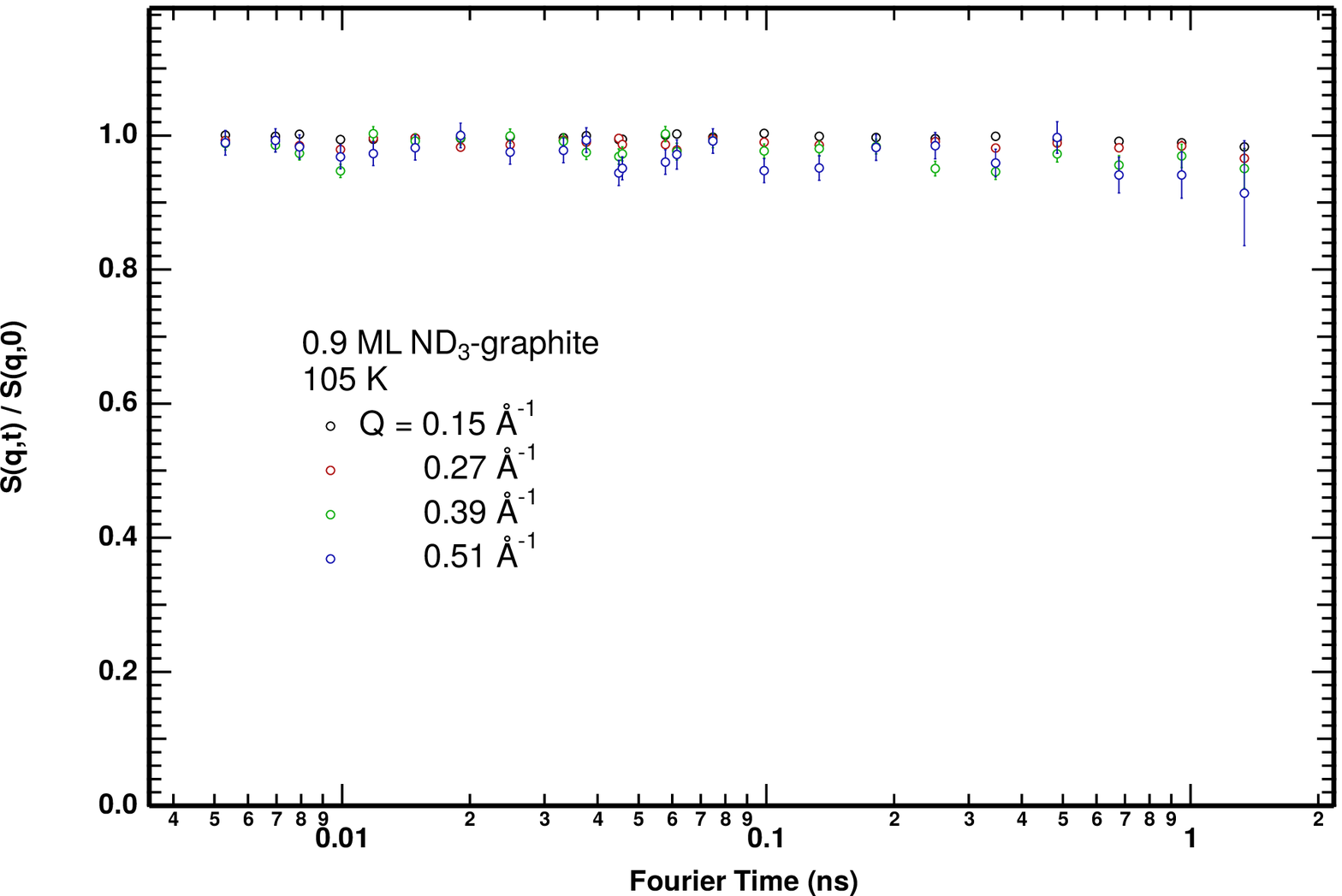}
     \caption{Neutron spin-echo spectra of 0.9 ML deuterated ammonia (ND$_3$) adsorbed on exfoliated graphite. The normalised intermediate scattering function $S(Q,t)/S(Q,0)$ shows hardly any change with Fourier time at a temperature of 105 K.}
     \label{fig:SEdata}
\end{figure}

As already mentioned in the main text, the quasi-elastic broadening determined from the TOF measurements corresponds to a broadening in time with $\tau \approx 1~\mbox{ps}^{-1}$ at $Q = 0.5~\mbox{\AA}^{-1}$. Diffusion at such a short timescale does not fit the current spectral window of IN11. As an example, \autoref{fig:SEdata} shows the normalised intermediate scattering function $S(Q,t)/S(Q,0)$ at 105 K for deuterated ammonia (ND$_3$) at a surface coverage of 0.9 ML. There appears no decay versus Fourier time within the given uncertainties. Only at the largest momentum transfer ($Q=0.51~\mbox{\AA}^{-1}$) one might anticipate a small change at about 1 ns. Hence the spin-echo measurements show that there is no additional motion at longer timescales, confirming the fast diffusion process seen in the TOF measurements.

\clearpage

\section{Fitting of the experimental data}
\begin{figure}[htb]
\centering
	\includegraphics[width = 0.95\textwidth]{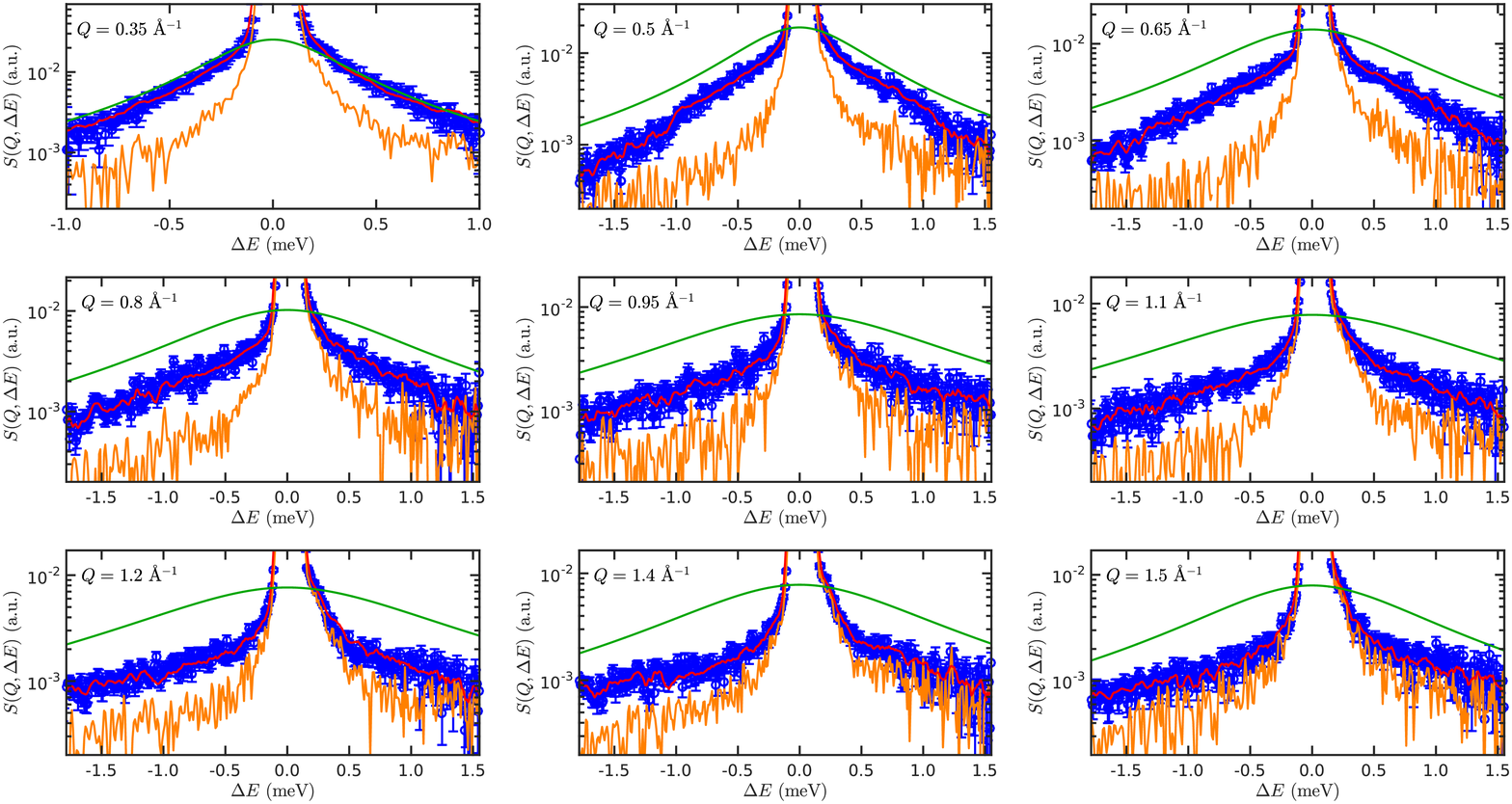}
\caption{Fitting of the experimentally measured $S(Q,\Delta E)$ (blue points) at 94 K. The orange curve is the resolution function (graphite sample measured at 4 K) and the green curve displays the single Lorentzian describing the quasi-elastic broadening. The red curve is the convolution according to \eqref{eq:scattering_function} of the resolution function with the quasi-elastic broadening and the elastic term.}
\label{fig:FitParam_94}
\end{figure}

\autoref{fig:FitParam_94} shows the result of the fit of the experimental data to \eqref{eq:scattering_function} (of the main paper) at 94 K and several momentum transfers. The red curve illustrates the convolution of the resolution function with the quasi-elastic broadening and the elastic term which is fitted to the experimental data. The orange curve is the resolution function, obtained by measuring the graphite sample measured at 4 K, and the green curve displays the single Lorentzian used to describe the quasi-elastic broadening.

\end{document}